\documentclass[reprint,superscriptaddress,amsmath,amssymb,aps,prl]{revtex4-1}

\usepackage{graphicx}
\usepackage{bm}
\usepackage[mathlines]{lineno}
\usepackage{mathtools}
\usepackage{natbib}
\usepackage[shortcuts]{extdash}
\usepackage{newtxtext}
\usepackage{color}

\usepackage{pdfpages}
\makeatletter\AtBeginDocument{\let\LS@rot\@undefined}\makeatother

\renewcommand{\vec}[1]{\bm{\mathrm{#1}}}
\newcommand{\vhat}[1]{\hat{\bm{\mathrm{#1}}}}

\let\epsilon\varepsilon

\begin{document}

\title{Unidirectional Magnon-Driven Domain Wall Motion \\ Due to the Interfacial Dzyaloshinskii-Moriya Interaction}

\author{Kyoung-Whan Kim}
\thanks{These two authors contributed equally to this work.}
\affiliation{Institute of Physics, Johannes Gutenberg University Mainz, 55099, Mainz, Germany}
\affiliation{Center for Spintronics, Korea Institute of Science and Technology, Seoul 02792, Korea}

\author{Seo-Won Lee}
\thanks{These two authors contributed equally to this work.}
\affiliation{Department of Materials Science and Engineering, Korea University, Seoul 02841, Korea}

\author{Jung-Hwan Moon}
\affiliation{Department of Materials Science and Engineering, Korea University, Seoul 02841, Korea}

\author{Gyungchoon Go}
\affiliation{Department of Materials Science and Engineering, Korea University, Seoul 02841, Korea}

\author{Aur\'{e}lien Manchon}
\affiliation{King Abdullah University of Science and Technology (KAUST), Physical Science and Engineering Division (PSE), Thuwal, 23955-6900, Saudi Arabia}
\affiliation{King Abdullah University of Science and Technology (KAUST), Computer, Electrical and Mathematical Science and Engineering Division (CEMSE), Thuwal, 23955-6900, Saudi Arabia}

\author{Hyun-Woo Lee}
\affiliation{Department of Physics, Pohang University of Science and Technology, Pohang 37673, Korea}

\author{Karin Everschor-Sitte}
\affiliation{Institute of Physics, Johannes Gutenberg University Mainz, 55099, Mainz, Germany}

\author{Kyung-Jin Lee}
\email{kj\_lee@korea.ac.kr}
\affiliation{Department of Materials Science and Engineering, Korea University, Seoul 02841, Korea}
\affiliation{KU-KIST Graduate School of Converging Science and Technology, Korea University, Seoul 02841, Korea}

\date{\today}

\begin{abstract}
We demonstrate a unidirectional motion of a quasiparticle without an explicit symmetry breaking along the space-time coordinate of the particle motion. 
This counterintuitive behavior originates from a combined action of two intrinsic asymmetries in the other two directions. We realize this idea with the magnon-driven motion of a magnetic domain wall in thin films with interfacial asymmetry. Contrary to previous studies, the domain wall moves along the same direction regardless of the magnon-flow direction. Our general symmetry analysis and numerical simulation reveal that the odd order contributions from the interfacial asymmetry is unidirectional, which is dominant over bidirectional contributions in the realistic regime. We develop a simple analytic theory on the unidirectional motion, which provides an insightful description of this counterintuitive phenomenon.
\end{abstract}

\pacs{ }

\maketitle

\paragraph{Introduction.---\hspace{-5pt}}

The motion of a physical particle is called unidirectional when it is along a particular direction (denoted by $x$) in spite of the presence of spatially symmetric excitations. The unidirectionality not only is physically interesting but also plays a central role in our real life as exemplified by diodes in electronic systems and molecular motors in biological systems~\cite{Vale1990}. Motivated by the Feynman ratchet~\cite{Feynman1966}, unidirectional motion is usually demonstrated in asymmetric potentials~\cite{Rousselet1994,Franken2012} or an energy gradient~\cite{Kou2012} along the motion direction, $x$. Unidirectional motion without spatial asymmetry has been suggested ~\cite{Cole2007}, but instead, it requires a time-asymmetric perturbation, i.e., a temporal ratchet. Therefore, the realization of the unidirectional motion has been limited to the cases where the symmetry is intentionally broken along the space-time coordinate of the particle motion ($x$ and $t$).

\begin{figure}
	\includegraphics[width=8.6cm]{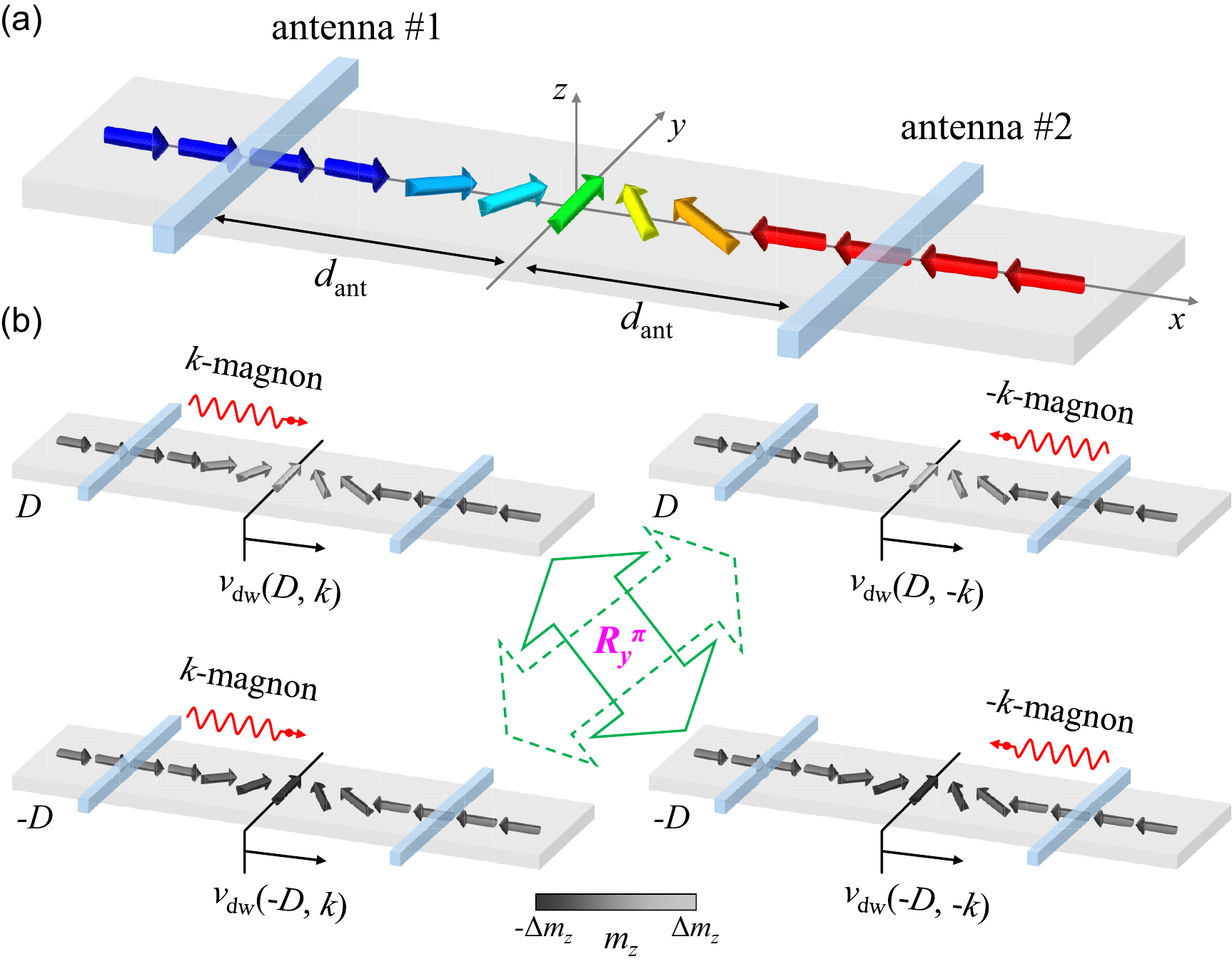}
	\caption{
		(a) One-dimensional magnetic system including an in-plane transverse DW at the center and two antennas with a distance $d_{\rm ant}$. The DMI is induced by inversion symmetry breaking along $z$. The antenna \#1 (\#2) generates a spin wave of momentum $+k$ ($-k$) toward the DW. 
		(b) Definition of the DW velocity $v_{\rm DW}(D,k)$ for $\pm$ signs of $D$ and $k$. The rotation of whole sample by $\pi$ around the $y$ axis (denoted by $R_y^\pi$) requires $v_{\rm DW}(D,k)=-v_{\rm DW}(-D,-k)$. The gray scale indicates a small tilting of the equilibrium DW structure by the DMI, which also flips the sign when $D$ changes its sign.
	}
	\label{fig:FIG1}
\end{figure}

In this Letter, we demonstrate that the explicit asymmetry along the space-time coordinate of the particle motion ($x$ and $t$) is not an essential condition for the unidirectional motion. The main idea is to exploit intrinsic asymmetries present in magnet-nonmagnet bilayers, i.e., the time reversal symmetry breaking of the magnetization and the structural inversion asymmetry of the bilayer, which make $x$ and $-x$ nonequivalent. Such broken symmetries are naturally realized in a magnetic system shown in Fig.~\ref{fig:FIG1}(a), where a magnetic domain wall (DW) particle possesses a controllable spontaneous asymmetry along $y$ (via the DW center magnetization in green) and the interface of the thin film provides an indispensable source of structural asymmetry along $z$~\cite{Kim2013}. The latter naturally generates the Dzyaloshinskii-Moriya interaction (DMI)~\cite{Dzialo1957,Moriya1960,Fert2013}, which is the antisymmetric component of the exchange interaction originating from spin-orbit coupling and broken inversion symmetry~\cite{Dzialo1957,Moriya1960,Fert2013}. The magnetic DW dynamics in the presence of the DMI has attracted considerable interest due to its rich physics and potential for applications~\cite{Thiaville2012,Emori2013,Ryu2013,Kim2012}.

We employ symmetry argument, micromagnetic simulation, and analytic theory to demonstrate unidirectional magnon-driven DW motion in systems with the above-mentioned intrinsic asymmetries, in contrast to previous theories that predict bidirectional magnon-driven DW motion regardless of its mechanism, such as the angular momentum transfer~\cite{Yan2011,Hinzke2011} and the linear momentum transfer~\cite{Han2009,Jamali2010,Seo2011,Wang2012,Yan2013,Wang2015,Yu2017}. Here the term ``unidirectional'' (``bidirectional'') refers to any contribution whose sign is independent of (dependent on) the sign of the external excitation (magnon injection direction in our case). 
We show that the unidirectional DW motion is generated not only by coherent spin waves but also by thermal magnons. Symmetric heating of both sides of the DW (but no heating at the DW position and thus not in thermal equilibrium) also induces the unidirectional motion, which would be experimentally testable. 
 Nevertheless, our work does not violate the fundamental laws of thermodynamics as the net DW velocity vanishes in thermal equilibrium. 

\paragraph{Symmetry argument.---\hspace{-5pt}} We make a symmetry argument for the unidirectionality of a quasiparticle motion in the presence of intrinsic symmetry breaking along $y$ and $z$. As an example, we consider an in-plane transverse magnetic DW in the presence of the DMI originating from an interface normal to $\vhat{z}$ (Fig.~\ref{fig:FIG1}). 
We define the magnon-induced DW velocity $v_{\rm DW}(D,k)$ as depicted in Fig.~\ref{fig:FIG1}(b), where $k$ is the incident spin-wave wave vector and $D$ is the strength of the interfacial DMI. The $k$ is positive (negative) when a spin wave is incident from the left (right) side of the DW. Depending on the signs of $D$ and $k$, there are four possible DW velocities; $v_{\rm DW}(\pm D,\pm k)$ and $v_{\rm DW}(\pm D,\mp k)$. These four velocities are related by a symmetry operation. When one rotates the whole system around the $y$ axis by $\pi$ (denoted by $R_y^\pi$), not only do the signs of $k$ and $v_{\rm DW}$ change, but also that of $D$ changes due to the reversal of the structural inversion asymmetry [Fig.~\ref{fig:FIG1}(b)]. This leads to the following general constraint:
\begin{subequations}
	\label{Eq:symmetry argument}
\begin{equation}
v_{\rm DW}(D,k)=-v_{\rm DW}(-D,-k).\label{Eq:symmetry argument-a}
\end{equation}
We now assume that $v_{\rm DW}(D,k)$ can be expanded perturbatively in $D$ (the validity is discussed below): $v_{\rm DW}(D,k)=v_0(k)+Dv_1(k)+D^2v_2(k)+\cdots$. Applying Eq.~(\ref{Eq:symmetry argument-a}) for each order of $D$ gives
\begin{equation}
v_n(k)=(-1)^{n+1}v_n(-k).\label{Eq:symmetry argument-b}
\end{equation}
\end{subequations}
Equation~(\ref{Eq:symmetry argument-b}) shows that 
the odd (even) order DMI contributions are unidirectional (bidirectional). For a symmetric excitation (i.e., simultaneous excitation of spin waves with $+k$ and $-k$), the bidirectional contributions are all canceled out; thus the net motion is unidirectional. Furthermore, if $|Dv_1|>|v_0|$ and the higher order contributions are negligible, $v_{\rm DW}(D,k)$ and $v_{\rm DW}(D,-k)$ have the same sign, giving a unidirectional motion even without asymmetric excitations. Note that our symmetry constraint [Eq.~(\ref{Eq:symmetry argument})] holds for any physical particle under arbitrary perturbations in films with (i) inversion symmetry breaking along $z$, (ii) the same boundary contribution under a symmetry operation ($R_y^\pi$), and (iii) higher order contributions of the asymmetry are negligible. 

There are two remarks. First, although the asymmetry along $y$ is not explicitly used for the symmetry argument, it is crucial for nonzero $v_{2n+1}(k)$. This is verified by taking $R_x^\pi$, implying $v_{\rm DW}(D,k)=v_{\rm DW}(-D,k)$ without an asymmetry along $y$. 
Second, our symmetry argument does not work for a DMI originating from bulk inversion asymmetry~\cite{Wang2015}, because its sign is not reversed under the rotation $R_y^\pi$ and, equivalently, it does not have an asymmetry along $z$.

\paragraph{Unidirectional DW motion driven by spin waves.---\hspace{-5pt}} We perform micromagnetic simulations to demonstrate the unidirectionality of the magnon-driven DW motion over wide ranges of parameters. The DW is initially positioned at the center of nanowire and spin-wave antennas [\#1 and \#2 in Fig.~\ref{fig:FIG1}(a)] are located $d_{\rm ant}$ distant from the initial DW position. A spin wave with $+k$ ($-k$) from antenna \#1 (antenna \#2) propagates toward the DW and gives rise to a DW displacement. 

We solve the Landau-Lifshitz-Gilbert equation, 
\begin{equation}
\partial_t\vec{m}=-\gamma\vec{m}\times\vec{H}_{\rm eff}+\alpha\vec{m}\times\partial_t\vec{m},
\end{equation}
where $\vec{m}$ is the unit vector along the magnetization, $\gamma$ is the gyromagnetic ratio, and $\alpha$ is the Gilbert damping constant. The effective field is given by 
\begin{equation}
{\mathbf H}_{\rm eff}
=\frac{2}{M_s}(A\partial_x^2\vec{m}-K_zm_z\vhat{z}+K_xm_x\vhat{x}-D\vhat{y}\times\partial_x\vec{m}),
\end{equation}
where $M_s$ is the saturation magnetization, $A$ is the exchange stiffness, $K_z=\mu_0 M_s^2/2$ is the hard-axis anisotropy, and $K_x$ is the easy-axis anisotropy. We take $D$ varying from 0.0~$\mathrm{mJ/m^2}$ to 1.0~$\mathrm{mJ/m^2}$, which is the typical order of magnitude considered in real systems with the interfacial DMI~\cite{Perez2014,Heinonen2016,Luchaire2016}. The simulation details including the materials parameters are presented in the Supplemental Material~\cite{Supple}.

Figure~\ref{fig:FIG2} shows computed DW velocity ($v_{\rm DW}$) induced by magnon with momentum $\pm k$. For $D=0$ [Fig.~\ref{fig:FIG2}(a)], $v_{\rm DW}$ is bidirectional and fits well with the previously reported velocity $v_{\rm DW}/|\rho|^2=-2k\gamma A/M_s$~\cite{Yan2011,Wang2012}, obtained from the angular momentum transfer mechanism without DMI, where $|\rho|^2$ is the injected magnon intensity. This corresponds to $v_0(k)$ in Eq.~(\ref{Eq:symmetry argument-b}). For $D=1.0~\mathrm{mJ/m^2}$ [Fig.~\ref{fig:FIG2}(b)], on the other hand, $v_{\rm DW}(+k)$ and $v_{\rm DW}(-k)$ have the same sign for whole tested ranges of $|k|$, demonstrating the DW unidirectionality. As $D$ increases from $0$ to $1.0~\mathrm{mJ/m^2}$, the unidirectionality first appears in high $k$ ranges and then expands to low $k$ ranges (not shown). For $D\geq0.5~\mathrm{mJ/m^2}$, the unidirectionality appears from a fairly low $k$ ($\approx 13\times10^5~\mathrm{cm^{-1}}$, corresponding wavelength $\approx 50~\mathrm{nm}$).

\begin{figure}
	\includegraphics[width=8.6cm]{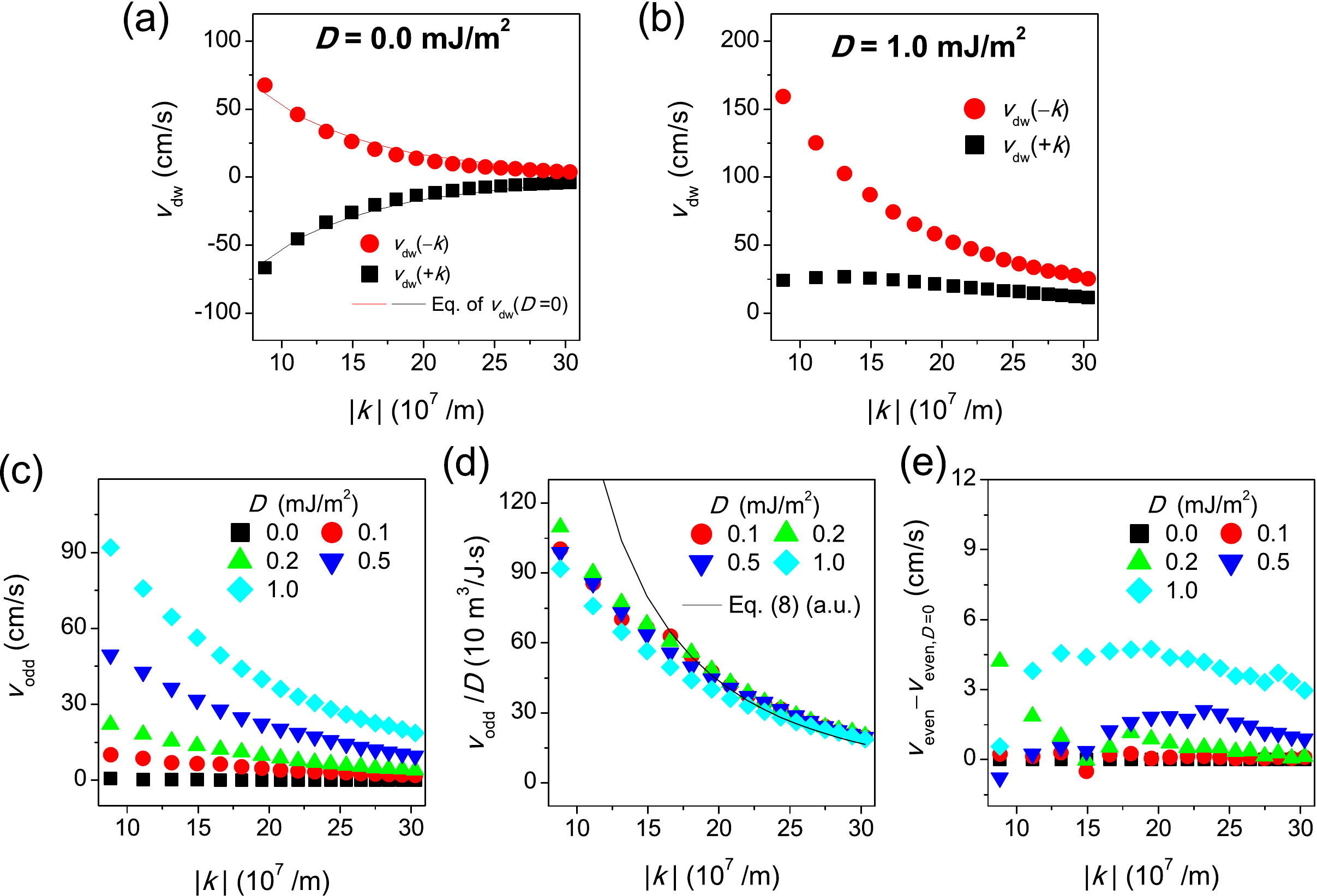}
	\caption{
		The magnon-driven DW velocity $v_{\rm DW}(\pm k)$ for (a) $D=0.0~\mathrm{mJ/m^2}$ and (b) $D=1.0~\mathrm{mJ/m^2}$ calculated for zero temperature.
		(c) Odd order contribution, $v_{\rm odd}=[v_{\rm DW}(+k)+v_{\rm DW}(-k)]/2$ for various $D$, which is comparable to or larger than the conventional velocity. 
		(d) Odd order contribution divided by $D$, which is almost independent of $D$, justifying the first order approximation. The solid line is calculated from Eq.~(\ref{Eq:final result}). (e) Even order DMI contributions, $v_{\rm even}=|v_{\rm DW}(+k)-v_{\rm DW}(-k)|/2-v_0$, implying that the higher order contributions are negligible.
	}
	\label{fig:FIG2}
\end{figure}

Figure~\ref{fig:FIG2}(c) shows the unidirectional contribution (odd order in $D$) calculated by $v_{\rm odd}=[v_{\rm DW}(D,k)+v_{\rm DW}(D,-k)]/2=Dv_1(k)+D^3v_3(k)+\cdots$. It clearly shows that for $D \geq 0.5~\mathrm{mJ/m^2}$, the unidirectional contribution is comparable to or larger than $v_0$ plotted in Fig.~\ref{fig:FIG2}(a) over a wide range of $k$. Despite the dominating DMI contribution to $v_{\rm DW}$, the perturbative expansion in Eq.~(\ref{Eq:symmetry argument-b}) is still valid. To justify this, we plot $v_{\rm odd}/D$ for various $D$ and show that the values are mostly independent of $D$ [Fig.~\ref{fig:FIG2}(d)]. Therefore, the unidirectional DW velocity is first order in $D$. Furthermore, we calculate $v_{\rm even}-v_0=D^2v_2(k)+D^3v_3(k)+\cdots$ for various $D$ to verify that the higher order contributions are negligible [Fig.~\ref{fig:FIG2}(e)]. The reason that the first order contribution $Dv_1$ can be larger than the zeroth order one $v_0$ is that they come from different physical origins: $v_0$ mainly originates from the angular momentum transfer mechanism~\cite{Yan2011} while $Dv_1$ mainly originates from the magnon-mediated Dzyaloshinskii-Moriya torque~\cite{Manchon2014}, as we show below.

\begin{figure}
	\includegraphics[width=8.6cm]{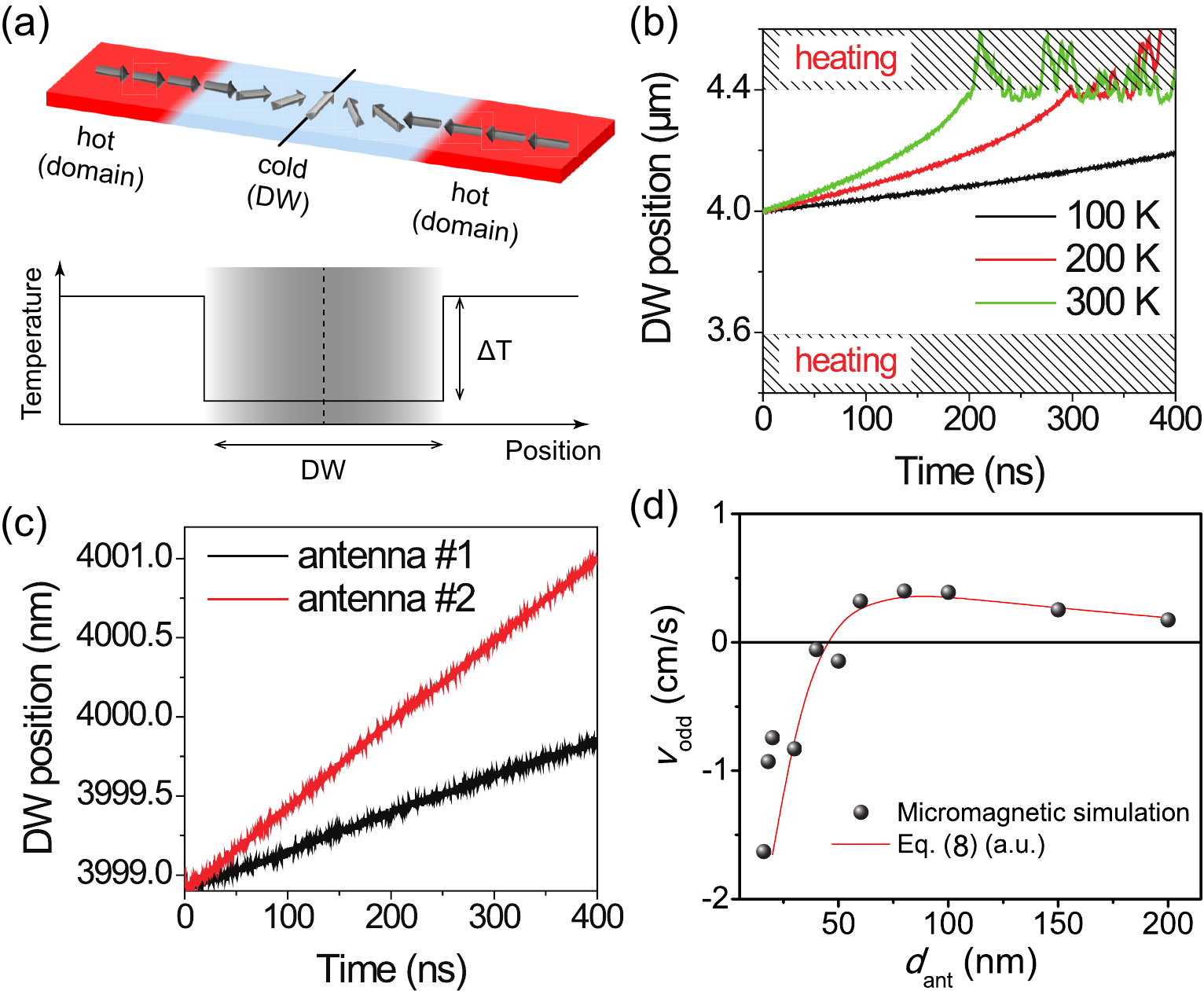}
	\caption{
		(a) Set up for symmetric heating of the domain parts. (b) The unidirectional motion under the symmetric heating. (c) The unidirectional DW motion due to thermal magnons from locally (two unit cells) heated antenna \#1 or \#2. Here a temperature of $300~\mathrm{K}$ is set at the antenna and $D=1.0~\mathrm{mJ/m^2}$ is used. The fluctuation of the data originates from the stochastic thermal fluctuation fields. (d) Local heating position dependence of DW velocity. The solid line is calculated from Eq.~(\ref{Eq:final result}).
	}
	\label{fig:FIG3}
\end{figure}

\paragraph{Unidirectional DW motion driven by thermal magnons.---\hspace{-5pt}} As a coherent spin wave with a single $k$ state induces a unidirectional DW motion in a wide range of $k$, thermal magnons consisting of many $k$ states are able to induce the DW unidirectionality. 
To demonstrate this, we heat up the domain parts to make them have a different temperature from that of the DW. Thus the system is in thermal nonequilibrium. Finite temperature effects are calculated by imposing the thermal fluctuation field~\cite{Brown1963,Supple} corresponding to the local temperature. We note that the temperature profile is \emph{symmetric} [Fig.~\ref{fig:FIG3}(a)]. Figure~\ref{fig:FIG3}(b) shows that the DW indeed moves towards a particular direction for various temperature differences $\Delta T$. The moving direction is determined by the DMI sign and the DW center magnetization direction~\footnote{To see the center magnetization dependence, one can take $R_z^\pi$ in Fig.~\ref{fig:FIG1}(a) to obtain the opposite $m_y$ and the opposite velocity.}: the latter is controllable by an external magnetic field. This offers a simple experimental scheme to observe the unidirectionality of the DW motion. In the experiment, the proposed \emph{symmetric}-heating setup will be useful to exclude the effect of temperature-dependent change in magnetic properties on the unidirectionality because they are also symmetric.

\begin{table}[b]
	\begin{tabular}{|c|c|c|c|}\hline
		&$T_{\rm DW}>T_{\rm domain}$&$T_{\rm DW}=T_{\rm domain}$&$T_{\rm DW}<T_{\rm domain}$\\\hline
		$Dm_{c,y}>0$&$v_{\rm DW}<0$ &$v_{\rm DW}=0$ & $v_{\rm DW}>0$\\\hline
		$Dm_{c,y}<0$&$v_{\rm DW}>0$ &$v_{\rm DW}=0$ &$v_{\rm DW}<0$ \\\hline
	\end{tabular}
	\caption{\label{Table:DW direction} The direction of the DW motion driven by symmetric heat. The sign of the direction of the DW motion is determined by the signs of $\Delta T=T_{\rm domain}-T_{\rm DW}$, the DMI, and the DW center magnetization ($m_{c,y}$).
			}
\end{table}

The result shown in Fig.~\ref{fig:FIG3}(b) suggests that a local heating of one of two antennas [depicted in Fig.~\ref{fig:FIG1}(a)] also generates a DW motion with different speeds depending on which antenna is heated up. This is verified by simulation results shown in Fig.~\ref{fig:FIG3}(c). We find that the DW moves towards a particular direction regardless of the direction of thermal magnon flow, proving that the velocity contribution summed up over various incoming thermal magnons is not canceled out. The resulting velocity is orders of magnitude smaller than that of Fig.~\ref{fig:FIG3}(b) because only two cells at the heated antenna position are heated. The observation in Fig,~\ref{fig:FIG3}(c), however, does not imply a finite velocity at thermal equilibrium. We observe from the simulation that an instantaneous DW velocity is random and thus the net velocity is zero when the whole system is subject to uniform temperature (not shown). This net zero velocity in thermal equilibrium can be understood as follows: when the DW part is also heated up at the temperature same as the domain parts, the thermal magnons \emph{outgoing} from the DW gives an opposite (negative) contribution to the DW velocity. Therefore, the net DW velocity at uniform temperature is canceled out as it appropriate in order to not violate the thermodynamic law. To verify this, we plot the unidirectional DW velocity as a function of the local heating position relative to the DW center. Figure~\ref{fig:FIG3}(d) shows that the DW velocity changes its sign: it is negative (positive) near (far away from) the DW. The large negative values near the DW center cancels the positive values far away from the DW, thus the total contribution is zero in thermal equilibrium. We summarize the DW motion direction with respect to the sign of $\Delta T$ in Table~\ref{Table:DW direction}.


\paragraph{Analytic theory.---\hspace{-5pt}} 
We develop an analytic theory to demonstrate the role of the dampinglike magnonic torque in the unidirectional DW motion. As justified in Fig.~\ref{fig:FIG2}(d), it suffices to develop a first order theory in $D$. We start from absorbing the DMI contribution in the effective field into the exchange field:
\begin{equation}
\frac{2}{M_s}(A\partial_x^2\vec{m}-D\vhat{y}\times\partial_x\vec{m})=\frac{2}{M_s}A\tilde{\partial}_x^2\vec{m}+\mathcal{O}(D^2),
\end{equation}
where $\tilde{\partial}_u$ is the chiral derivative~\cite{Kim2013}, defined by
\begin{subequations}
\begin{align}
\tilde{\partial}_u\vec{f}&=\partial_u\vec{f}-\frac{D}{2A}(\vhat{z}\times\vhat{u})\times\vec{f},\\
\tilde{\partial}_uf&=\partial_uf,
\end{align}
\end{subequations}
for an arbitrary vector function $\vec{f}$ and scalar function $f$. Thus any DMI contribution can be obtained by replacing ordinary derivatives by the chiral derivatives in previous theories~\cite{Kovalev2012,Kim2015a} developed without considering the DMI~\cite{Supple}.

We use the previous theory on magnonic torque without the DMI~\cite{Kim2015a};
\begin{equation}
\vec{\tau}_{ D=0}=\hbar J_0\partial_x\vec{m}_0-(A\partial_x|\rho|^2)\vec{m}_0\times\partial_x\vec{m}_0,\label{Eq:magnonic torque no DMI}
\end{equation}
where $J_0=(2A/\hbar)[\vec{m}_0\cdot\langle\delta\vec{m}\times\partial_x\delta\vec{m}\rangle]$ is the magnon-flux density evaluated in the absence of the DW, $\vec{m}_0$ is the equilibrium DW profile, $\delta\vec{m}=\vec{m}-\vec{m}_0$ is the magnon excitations, $|\rho|^2=\langle\delta\vec{m}^2\rangle$ is proportional to the magnon number density, and $\langle\cdots\rangle$ is the time average over the (rapid) spin-wave fluctuation, thus $\vec{\tau}_{D=0}$ is a torque acting on slow DW dynamics. Replacing $\partial_x$ by $\tilde{\partial}_x$ in Eq.~(\ref{Eq:magnonic torque no DMI}) gives the DMI corrections:
\begin{align}
\vec{\tau}&=\hbar J\partial_x\vec{m}_0-(A\partial_x|\rho|^2)\vec{m}_0\times\partial_x\vec{m}_0\nonumber\\
&\quad-\frac{D\hbar}{2A} J_x(\vhat{y}\times\vec{m}_0)+\frac{D}{2}(\partial_x|\rho|^2)\vec{m}_0\times(\vhat{y}\times\vec{m}_0),\label{Eq:magnonic torque}
\end{align}
where $J=(2A/\hbar)[\vec{m}_0\cdot\langle\delta\vec{m}\times\tilde{\partial}_x\delta\vec{m}\rangle]$ is the modified magnon-flux density due to DMI-induced change in the magnon dispersion. The first two terms in Eq.~(\ref{Eq:magnonic torque}) are the adiabatic~\cite{Yan2011,Wang2012} and nonadiabatic magnonic torques~\cite{Kim2015a,Kovalev2014} respectively, and the third and fourth terms are fieldlike and dampinglike Dzyaloshinskii-Moriya torques~\cite{Manchon2014}, respectively.

To obtain the DW velocity, we use $v_{\rm DW}\propto\int (dm_x/dt)dx\propto\int \tau_x dx$. The second and third terms in Eq.~(\ref{Eq:magnonic torque}) do not contribute to $v_{\rm DW}$ because of the parity of $\vec{m}_0$. As a result, we obtain
\begin{equation}
v_{\rm DW}\propto\int \hbar J\partial_xm_{0,x}dx+\int\frac{D}{2}(\partial_x|\rho|^2)m_{0,x}m_{0,y}dx. \label{Eq:final result}
\end{equation}
The first term is the conventional angular momentum transfer contribution~\cite{Yan2011} which is bidirectional.
The second term is the dampinglike magnonic torque contribution which is unidirectional. To see the unidirectionality, one takes the inversion of the integrand ($x\to-x$) to obtain $\partial_x|\rho|^2\to-\partial_x|\rho|^2$ and $m_{0,x}m_{0,y}\to-m_{0,x}m_{0,y}$, thus the contribution does not change its sign upon the inversion. From Eq.~(\ref{Eq:final result}), one finds that in thermal equilibrium (uniform temperature), $J=\partial_x|\rho|^2=0$ implies the absence of the DW velocity. Equation~(\ref{Eq:final result}) is used to obtain the solid lines in Figs.~\ref{fig:FIG2}(d) and \ref{fig:FIG3}(d)~\cite{Supple}. For Fig.~\ref{fig:FIG2}(d), our first-order theory gives reasonable unidirectional DW velocities for large $k$, but some deviations for small $k$. The deviations may originate from the breakdown of the continuum model for the scattering of magnons by a DW, which has been shown even without the DMI~\cite{Wang2012}. For Fig.~\ref{fig:FIG3}(d), on the other hand, Eq.~(\ref{Eq:final result}) describes the numerical results well, justifying the validity of our first-order theory. For more motivated readers, we present in the Supplemental Material~\cite{Supple} more remarks on our formalism, which will be useful for generalizing our result.


\paragraph{Discussion.---\hspace{-5pt}} We demonstrate a unidirectional magnon-induced DW motion in the presence of the interfacial DMI. Unlike previously demonstrated unidirectional motions, our theory does not require an explicit asymmetry along $x$ and $t$, but exploits intrinsic asymmetries present along $y$ and $z$. Therefore, our work sheds light on the mechanism of unidirectionality by demonstrating that an explicit asymmetry along the space-time coordinate of the particle motion is not essential for realizing the particle unidirectionality.

As we use the asymmetry intrinsically present in the system, on the other hand, our work is intimately related to the ongoing researches on the nonreciprocal response~\cite{Tokura2018}, which is referred to as directional transport and propagation of microscopic quantum particles such as electron, photon, magnon, and phonon, and is known to be present in materials system with broken inversion symmetry. A distinct difference of our work is that the nonreciprocal response appears even for a macroscopic classical particle, i.e., a magnetic DW. In this respect, our work will contribute to expand the research scope of the nonreciprocal response to macroscopic classical particles. We believe this contribution is important as classical particles are easy to manipulate and detect, thereby offering a framework to investigate the nonreciprocal response in wider contexts than examined before.

\paragraph{Note added.---\hspace{-5pt}} Not long ago, we became aware of a recent work~\cite{Psaroudaki2018} predicting a unidirectional motion of a Skyrmion under an oscillatory magnetic field.

\begin{acknowledgments}
We acknowledge A.~Thiaville, P.~Pirro, and S.-K.~Kim for fruitful discussions. This work is supported by NRF (2015M3D1A1070465, 2017R1A2B2006119) and the KIST Institutional Program (No. 2V05750 and No. 2E29410). The work in Mainz was supported by the Alexander von Humboldt Foundation, the ERC Synergy Grant SC2 (No.~610115), the Transregional Collaborative Research Center (SFB/TRR) 173 SPIN+X, and the German Research Foundation (DFG) (No. EV 196/2-1 and No. SI 1720/2-1). K.W.K was also supported by the National Research Council of Science \& Technology (NST) (Grant No. CAP-16-01-KIST) funded by the Korea government (Ministry of Science and ICT). H.W.L. was supported by NRF (2018R1A5A6075964). A.M. acknowledges support from the King Abdullah University of Science and Technology (KAUST).
\end{acknowledgments}



\section*{Supplementary material}

# Supplemental Material for
# "Unidirectional Magnon-Driven Domain Wall Motion Due to the Interfacial Dzyaloshinskii-Moriya Interaction"

Kyoung-Whan Kim, Seo-Won Lee, Jung-Hwan Moon, Gyungchoon Go,
Aurélien Manchon, Hyun-Woo Lee, Karin Everschor-Sitte, and Kyung-Jin Lee*

## I. DETAILS FOR MICROMAGNETIC SIMULATIONS

### A. DW motion driven by spin waves

To solve the Landau-Lifshitz-Gilbert equation [Eq. (2)] numerically, we discretize the system along $x$ with the unit length $a = 2$ nm. We employ a spin-wave-absorbing boundary condition by increasing $\alpha$ smoothly near edge (500 unit cells at both sides) to $\alpha_{\max} = 1.00$ for preventing spin-wave reflection [S1]. To excite spin waves with frequency $f$, we apply an ac magnetic field $H_{\rm ac}\sin(2\pi ft)$ on two unit cells at the locations of the antennas, where $H_{\rm ac} = 1200$ Oe. The DW velocity is calculated by the ratio between the simulation time and the change of the DW position where $m_x$ vanishes.

The modeling parameters are: the total length of nanowire $L = 8$ μm, the distance between the DW and the antenna $d_{\rm ant} = 600$ nm, $M_s = 800$ kA/m, $A = 13$ pJ/m, $K_x = 47$ kJ/m$^3$, and $\alpha = 0.01$. In this system, the equilibrium DW satisfies the Walker profile $\mathbf{m}_0 = (-\tanh(x-X)/\lambda, \mathrm{sech}(x-X)/\lambda, 0)$ where $\lambda = \sqrt{A/K_x} \approx 16.6$ nm is the DW width and $X$ is the DW position.

To plot Fig. 2 in the main text, the dispersion relation $f = (\gamma/\pi M_s)\sqrt{(K_x + Ak^2)(K_z + K_x + Ak^2)}$, which is also plotted in Fig. S1, is used.

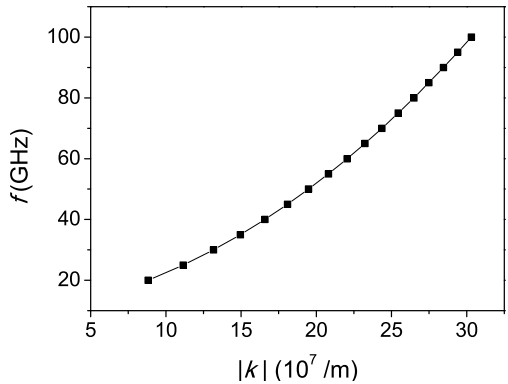

FIG. S1. The dispersion relation of generated spin waves.

---

* kj_lee@korea.ac.kr

### B. DW motion driven by thermal magnons

For simulations on the thermal effects, we superimpose the thermal fluctuation field $\mathbf{H}_{\rm th}$ onto $\mathbf{H}_{\rm eff}$ [Eq. (3)] and solve the *stochastic* Landau-Lifshitz-Gilbert equation. The thermal fluctuation field is random $\langle \mathbf{H}_{\rm th} \rangle = 0$ and satisfies the fluctuation-dissipation theorem $\langle H_{{\rm th},i}(x',t')H_{{\rm th},j}(x,t)\rangle = (2\alpha k_B T/\gamma M_s S)\delta_{ij}\delta(x-x')\delta(t-t')$ [S2], where $k_B$ is the Boltzmann constant, $S$ is the cross sectional area, and $\langle \cdots \rangle$ is the ensemble average. We generate random numbers for every step to simulate the thermal fluctuation. Under the thermal fluctuation field, the principles of statistical physics automatically excite a bunch of magnon modes following the Bose-Einstein distribution, which we call thermal magnons in the main text.

## II. DETAILS FOR ANALYTIC THEORY

### A. Mathematical justification behind the chiral derivative replacement

First we briefly review the derivation of Eq. (6), which is derived in Ref. [S3]. They start from the Landau-Lifshitz-Gilbert equation with the effective magnetic field in the form of $\mathbf{H}_{\rm eff} = (2A/M_s)\sum_i \partial_i^2 \mathbf{m} - \sum_{ij} K_{ij} m_i \hat{\mathbf{j}}$ and decompose the magnetization into a slowly varying contribution (such as the DW motion) and a rapidly varying contribution (the spin wave oscillation). The time scales of these two components are assumed to be well separated. After taking the time average of the fast degree of freedom, one obtains the equation of motion for the slow degree of freedom driven by the fast degree of freedom. This procedure gives the effective magnon-driven spin torque. We note that, in the derivation there, the following basic properties of the derivatives are used.

$$\partial_u(\mathbf{m} + \mathbf{n}) = \partial_u \mathbf{m} + \partial_u \mathbf{n}, \tag{S1a}$$
$$\partial_u(f + g) = \partial_u f + \partial_u g, \tag{S1b}$$
$$\partial_u(f\mathbf{m}) = f\partial_u \mathbf{m} + \mathbf{m}\partial_u f, \tag{S1c}$$
$$\partial_u(\mathbf{m} \cdot \mathbf{n}) = \mathbf{m} \cdot \partial_u \mathbf{n} + \mathbf{n} \cdot \partial_u \mathbf{m}, \tag{S1d}$$
$$\partial_u(\mathbf{m} \times \mathbf{n}) = \mathbf{m} \times \partial_u \mathbf{n} + \partial_u \mathbf{m} \times \mathbf{n}, \tag{S1e}$$
$$\mathbf{m} \cdot \partial_u \mathbf{m} = 0. \tag{S1f}$$

However, the other properties of the derivative or an explicit definition of $\partial_u$ is unnecessary in the mathematical derivation.

Therefore, one can straightforwardly generalize the previous formalism by the following way. For any linear operator $\mathcal{L}_i$ satisfying Eq. (S1) (after replacing $\partial_u$ by $\mathcal{L}_u$), the effective

field $\mathbf{H}_{\text{eff}} = (2A/M_s) \sum_i \mathcal{L}_i^2 \mathbf{m} - \sum_{ij} K_{ij} m_i \hat{\mathbf{j}}$ (as assumed in Ref. [S3]) gives exactly the same result at *every* algebraic step of the derivation. Therefore, the resulting magnon-induced spin torque is given by the same expression except the replacement $\partial_u \to \mathcal{L}_u$. In our case [Eq. (4)], the effective field is given by the chiral derivative $\mathcal{L}_x = \tilde{\partial}_x$. Remarkably, from the definitions of the chiral derivative [Eq. (5)], one can verify the following algebraic rules.

$$\tilde{\partial}_u(\mathbf{m} + \mathbf{n}) = \tilde{\partial}_u \mathbf{m} + \tilde{\partial}_u \mathbf{n}, \quad \text{(S2a)}$$

$$\tilde{\partial}_u(f + g) = \tilde{\partial}_u f + \tilde{\partial}_u g, \quad \text{(S2b)}$$

$$\tilde{\partial}_u(f\mathbf{m}) = f\tilde{\partial}_u \mathbf{m} + \mathbf{m}\tilde{\partial}_u f, \quad \text{(S2c)}$$

$$\tilde{\partial}_u(\mathbf{m} \cdot \mathbf{n}) = \mathbf{m} \cdot \tilde{\partial}_u \mathbf{n} + \mathbf{n} \cdot \tilde{\partial}_u \mathbf{m}, \quad \text{(S2d)}$$

$$\tilde{\partial}_u(\mathbf{m} \times \mathbf{n}) = \mathbf{m} \times \tilde{\partial}_u \mathbf{n} + \tilde{\partial}_u \mathbf{m} \times \mathbf{n}, \quad \text{(S2e)}$$

$$\mathbf{m} \cdot \tilde{\partial}_u \mathbf{m} = 0, \quad \text{(S2f)}$$

each of which corresponds to that of Eq. (S1). Therefore, Eq. (S2) implies that one may elegantly derive the magnon-induced spin torque [Eq. (7)] by simply replacing the ordinary derivative by the chiral derivative.

### B. Fitting procedure

The analytic theory [Eq. (8)] gives the following unidirectional contribution to the DW velocity,

$$v_{\text{odd}} \propto \int \frac{D}{2}(\partial_x |\rho|^2) m_{0,x} m_{0,y} dx. \quad \text{(S3)}$$

To obtain the fitting curves in the main text, we impose the profile of $|\rho|^2 = \rho_0 e^{-|x-x_0|/\lambda_m}$, where $\lambda_m$ is the magnon decay length and $x_0$ is the local heating position (*i.e.*, the antenna). To determine $\rho_0$ and $\lambda_m$, we run a separate simulation without the DW, since the correction due to the DW gives a higher order contribution in $D$. The obtained magnon decay length is $\lambda_m = 150$ nm for the thermal magnons [Fig. 3(d)] and is $k$ dependent for the spin wave excitation [Fig. 2(d)]. The proportionality constant in Eq. (S3) is also taken to be a fitting parameter. Together with $|\rho|^2$ obtained by this procedure, the Walker profile for $\mathbf{m}_0$ (See Sec. I A) gives the solid lines in the main text.

### III. MORE ON THE DM TORQUE

#### A. Unidirectionality understood by visualization

The unidirectionality from the dampinglike DM torque can be understood graphically. In Fig. S2(a) [Fig. S2(b)], we draw $\partial_x|\rho|^2 \mathbf{m}_0 \times (\hat{\mathbf{y}} \times \mathbf{m}_0)$ for a positive bias $\partial_x|\rho|^2 > 0$ (negative bias $\partial_x|\rho|^2 < 0$). Based on the area we highlight, from which the dominant contribution comes, the illustration clearly shows that the DW moves right (for a positive $D$), regardless of the magnonic bias direction.

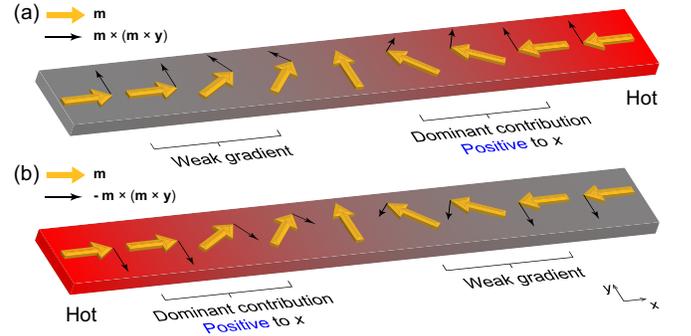

FIG. S2. Illustration of the unidirectionality originating from the dampinglike DM torque. (a) Under a positive magnonic bias. In the grayed area, the magnonic bias exponentially decays, thus the red part gives the dominant contribution, where the magnonic torque is along the positive $x$ direction. (b) Under a negative bias, the dominant contribution comes from the opposite side where the reversed magnonic torque is again along the positive $x$ direction. This qualitatively explains the unidirectionality shown in Figs. 2(b) and 3(c).

#### B. Understanding by integration by parts

Regarding Eq. (8), in addition to the discussion in the main text, another intuitive understanding is achieved by integrating Eq. (S3) by parts.

$$v_{\text{odd}} \propto -\int \frac{D}{2}|\rho|^2 \partial_x(m_{0,x} m_{0,y}) dx. \quad \text{(S4)}$$

For simulations in Figs. 2, 3(c), and 3(d), one may consider $|\rho|^2$ being localized at the excited antenna.

Equation (S4) provides another way to understand the unidirectionality: from the Walker profile for $\mathbf{m}_0$ (See Sec. I A), one concludes that $\partial_x(m_{0,x} m_{0,y})$ is an even function of $x$ [S4]. The absence of the DW motion at thermal equilibrium can also be easily understood by taking $|\rho|^2$ to be constant and deducing $v_{\text{odd}} = 0$. In addition, one can easily understand the sign change in Fig. 3(d), by noting that $-\partial_x(m_{0,x} m_{0,y})$ in the integrand gives a large negative contribution when $|\rho|^2$ is localized near the DW while it gives a positive contribution when $|\rho|^2$ is localized far away from the DW.

#### C. Generalization of the DMI chiral derivative

Our theoretical formalism can easily be generalized for systems with other symmetry by using the generalized chiral derivatives, which was introduced for spin-orbit coupled systems [S5]. Here we present the DMI version of the generalized chiral derivative for an arbitrary antisymmetric exchange interaction.

Starting from the energy density of the generalized DMI $\varepsilon_D = \sum_{ijk} D_{ijk} m_i \partial_j m_k$ with an antisymmetric DMI tensor ($D_{kji} = -D_{ijk}$) [S6], one obtains the generalized DMI field given by $\mathbf{H}_{\text{eff,DMI},i} = -(2/M_s) \sum_{jk} D_{ijk} \partial_j m_k$. Adding this to the exchange field is gives the generalized DMI chiral derivative in the following form.

$$\frac{2}{M_s}\left(A\nabla^2\mathbf{m} - \sum_{ijk}\hat{\mathbf{i}}D_{ijk}\partial_j m_k\right) = \frac{2}{M_s}A\tilde{\nabla}^2\mathbf{m} + \mathcal{O}(D^2), \quad \text{(S5)}$$

where $\hat{\mathbf{i}}$ is the unit vector along the $i$ direction and the generalized DMI chiral derivative is

$$\tilde{\partial}_u\mathbf{f} = \partial_u\mathbf{f} - \frac{1}{2A}\sum_{ij}\hat{\mathbf{i}}D_{iuj}f_j, \quad \text{(S6a)}$$

$$\tilde{\partial}_u f = \partial_u f. \quad \text{(S6b)}$$

We note that the generalized chiral derivative also satisfies Eq. (S2), thus our theory in the main text can be straightforwardly generalized.



\end{document}